\documentclass[aps,prl,amsmath,amssymb,amsfonts,twocolumn,showpacs]{revtex4-1}
\usepackage[cp1251]{inputenc}
\usepackage[russian]{babel}
\usepackage{graphicx}

\DeclareMathOperator{\tr}{tr}

\begin{document}

\title{On the macroscopic quantization in mesoscopic rings and single-electron devices}

\author{Andrew G. Semenov
}
\email{semenov@lpi.ru}
\affiliation{I.E.Tamm Department of Theoretical Physics, P.N.Lebedev
Physics Institute, 119991 Moscow, Russia
}

\begin{abstract}
In this letter the phenomenon of macroscopic quantization  is investigated using the particle on the ring interacting with the dissipative environment as an example. It is  shown that the phenomenon of macroscopic quantization has the clear physical origin in that case. It follows from the angular momentum conservation combined with momentum quantization for bare particle on the ring . The existence an observable which  can take only integer values in the zero temperature limit is rigorously proved. With the aid of the mapping between particle on the ring and Ambegaokar-Eckern-Schon model, which can be used to describe single-electron devices, it is  demonstrated that this observable is analogous to  the ``effective charge'' introduced by Burmistrov and Pruisken for the single-electron box problem. Different consequences of the  revealed physics are discussed, as well as a generalization of the obtained results  to the case of more complicated systems. 
\end{abstract}

\maketitle

\paragraph{Introduction.}
The phenomenon of Coulomb blockade \cite{ShZ1990,GZ1992,GZ1994,KSS1995,SS1994,GKSShZ1997} in different meso- and nanostructures is one of the most 
striking  manifestation of charge quantization. For example it can be observed in the so called single-electron box. This system consists of metallic island coupled by the tunnel junction to the electrode and capacitively coupled to the gate voltage. Varying the gate voltage one can change the number of electrons on the island and if the temperature much lower than the Coulomb energy of the grain, its average charge  will be the integer number of elementary electron charges $e$ . Theoretically such a system can be described by well-known Ambegaokar-Eckern-Schon (AES) approach \cite{AES1982,AES1984,ShZ1990} employing the smallness of tunnel barrier transparency. It is well known, that one can interpret  AES action as  an effective action for the particle on the ring interacting with the dissipative environment \cite{Guinea2002,Guinea2003,GHZ2003,EBD2012,SZ2011}. From the form of the action it follows that the environment is of Caldeira-Leggett (CL) type \cite{CL1981,ShZ1990},  but in principle other types of linear 
dissipative environments possible. If we put the ring into the magnetic field,  
persistent current 
will flow around it, so such a system can be used for the investigation of the influence of interactions on the coherent phenomena such as persistent current, its fluctuations and so on. Also this mapping has been extensively used by various authors in the numerical computations of the single-electron box properties (see for example \cite{HSZ1999,GHZ2003} and references therein).

As we mention above the average charge on the island is quantized but strictly speaking this statement is true only in the limit of zero temperature and vanishing tunnel coupling between the island and electrode. In general case the average charge is continuous. Recently it was argued by Burmistrov and Pruisken  \cite{BP2008,BP2010} (see also Ref. \cite{Bulgadaev2006}) that in such a system there is another quantity, ``effective charge'' which is nevertheless quantized in the limit of zero temperature. They relate such a quantity to the sensitivity of the single-electron device to the changing of boundary conditions and employ the similarity between AES theory and the theory of quantum Hall effect to introduce the unifying scaling diagram of a problem. Also it was supported  by explicit calculations in the cases of the small and large coupling constant. Note, that intimate connection between the quantum Hall effect and Coulomb blockade was previously revealed and investigated in the works \cite{Apenko2006,Apenko2008,
Bulgadaev2006}. But what is the counterpart of the Burmistrov and  Pruisken ``effective charge''  in the problem of the particle on the ring and does the phenomenon of macroscopic quantization  present in the case of other dissipative environments different from the CL one. The aim of the present letter is to prove rigorously the existence of macroscopically 
quantizing observable  in presence of different linear environments and to show that on  the language of particle on the ring it is deeply related to the total angular momentum conservation for the combined system particle plus the environment. 

The paper organized as followed. At the beginning we introduce the partition function for the  particle on the ring interacting with the environment and its relation to the single-electron box problem. After that  we rewrite it in the operator formalism and introduce fictitious many-particle system. On the next step we present a conserving quantity which commutes with Hamiltonian and has integer eigenvalues. At the and of the letter we relate it to the total angular momentum of the whole system and show, that it has nontrivial form in the original single-particle variables.

\paragraph{Model and basic definitions.}

Let us consider particle on the ring interacting with the linear dissipative environment at temperature $T$. It is well known that partition function can be represented through the path integral as \cite{Guinea2002,Guinea2003,GHZ2003}
\begin{equation}
\mathcal Z=\int\limits_{0}^{2\pi}d\theta_0\sum\limits_{n=-\infty}^\infty e^{2\pi i n\phi_x}\int\limits_{\theta(0)=\theta_0}^{\theta(\beta)=\theta_0+2\pi n}\mathcal D\theta e^{-S[\theta(\tau)]}
\label{partf}
\end{equation}
where
\begin{multline}
 S[\theta]\equiv S_0[\theta]+S_{i}[\theta]=\frac{mR^2}{2}\int\limits_0^\beta\dot\theta^2(\tau)d\tau\\+\frac{g}{4}\int\limits_0^\beta d\tau\int\limits_0^\beta d\tau'\alpha(\tau-\tau')e^{i\theta(\tau)-i\theta(\tau')}. 
 \label{effact}
\end{multline}
Here $m$ and $R$ are  the mass of the particle and ring radius, $g$ is the coupling constant between particle and environment, $\beta=1/T$ is inverse temperature, and kernel $\alpha(\tau-\tau')$ is governed by statistical and dynamical properties of the environment. Integration is performed over trajectories with a given winding number $n$, which is periodic up to constant $\theta(\beta)=\theta(0)+2\pi n$. $\phi_x=\Phi/\Phi_0$ is the flux piercing the ring in the units of flux quanta $\Phi_0=2\pi /e$ (here and below we set the Planck’s constant and the speed of light equal to unity $\hbar=1,\ c=1$). Due to bosonic nature of the environment, kernel $\alpha(\tau)$ is the periodic function of imaginary time $\tau$ with period equals to $\beta$ means that it can be represented in the form
\begin{equation}
 \alpha(\tau)=\frac{T}{\pi}\sum\limits_{n=-\infty}^\infty F(\omega_n)e^{-i\omega_n\tau},
\end{equation}
where $\omega_n=2\pi n T$ is Matsubara frequency and function $F(z)$ is symmetric $F(z)=F(-z)$ and equals to zero at zero frequency $F(0)=0$. Usually, the last condition might be achieved by substraction of some unimportant constant from the initial action. The CL environment \cite{ShZ1990,CL1981} corresponds to $F(z)=|z|$. This case is very important  from the practical point of view, since it is equivalent  to the AES model describing the single-electron box.  In order to relate these two problems one should identify $1/(2m R^2)$ with charging energy $E_c$  of the island, $g$ with dimensionless  conductance $g_t$ of the tunnel junction between island and reservoir, and flux $\phi_x$ with the  external charge $q_x$ induced by gate voltage. 

\paragraph{Decoupling of the action.} The key idea of our calculations is to transform initial problem with partition function given by equation (\ref{partf}) back to operator formalism and to introduce fictitious many-particle system interacting with the particle on the ring. We consider the following many-particle Hamiltonian
\begin{multline}
\hat H_f=\frac{(\hat p-\phi_x)^2}{2mR^2}+\sum\limits_k \varepsilon_k\left(\hat a^\dag_k\hat a_k+\hat b^\dag_k\hat b_k\right)\\+\sqrt{g}e^{i\hat\theta}\sum_k(\hat a_k+\hat b^\dag_k)
+\sqrt{g}e^{-i\hat\theta}\sum_k(\hat a_k^\dag+\hat b_k).
\label{ficts}
\end{multline}
Here $\hat \theta$ is  the angle variable  corresponded to the position of the particle on the ring, and $\hat p$ is the angular momentum operator which is conjugate to the position operator $[\hat p,\hat \theta]=-i$. In the position representation it acts as derivative $\hat p=-i\frac{\partial}{\partial\theta}$. Also fictitious system contains two sets of bosonic modes. It creation and annihilation operators are denoted by $\hat a^\dag_m,\hat b^\dag_m$ and $\hat a_m,\hat b_m$ correspondingly. Commutation relations are standard
\begin{equation}
[\hat a_k,\hat a_l]=[\hat a_k,\hat b_l]=[\hat b_k,\hat b_l]=[\hat a_k,\hat b^\dag_l]=0,
\end{equation}
\begin{equation}
[\hat a_k,\hat a_l^\dag]=[\hat b_k,\hat b_l^\dag]=\delta_{k,l},
\end{equation}
where $\delta_{k,l}$ is the Kroenecker symbol. Let us connect our fictitious system to the initial problem. This can be done by standard transformation of the partition function $\mathcal Z=\tr{e^{-\beta\hat H_f}}$ into  path integral representation and subsequent integration over bosonic variables. This procedure leads to the effective action as in equation (\ref{effact}) with
\begin{equation}
\alpha(\tau)=-4\sum\limits_k\langle\mathcal T(\hat a_k(\tau)+\hat b^\dag_k(\tau))(\hat a_k^\dag(0)+\hat b_k(0))\rangle_{a,b}.
\end{equation}
Here averaging is performed with density matrix of bath decoupled from particle $ \langle...\rangle_{a,b}\equiv\tr(...e^{-\beta\hat H_b})/\tr(e^{-\beta\hat H_b})$, where bath Hamiltonian is 
\begin{equation}
\hat H_b=\sum\limits_k \varepsilon_k\left(\hat a^\dag_k\hat a_k+\hat b^\dag_k\hat b_k\right),
\end{equation}
$\mathcal T$ is the time ordering symbol, and 
\begin{eqnarray}
\hat a_k(\tau)=e^{\tau\hat H_b}\hat a_k e^{-\tau\hat H_b}&\qquad& \hat a_k^\dag(\tau)=e^{\tau\hat H_b}\hat a_k^\dag e^{-\tau\hat H_b}\\
\hat b_k(\tau)=e^{\tau\hat H_b}\hat b_k e^{-\tau\hat H_b}&\qquad& \hat b_k^\dag(\tau)=e^{\tau\hat H_b}\hat b_k^\dag e^{-\tau\hat H_b}
\end{eqnarray} 
are operators in Matsubara representation. By averaging over bath variables one can obtain the desired relation
\begin{multline}
F(\omega_n)=-8\pi\sum_k\frac{\varepsilon_k}{\omega_n^2+\varepsilon_k^2}+C\\=-8\pi\int\limits_0^\infty \frac{dz}{2\pi} \frac{zJ(z)}{\omega_n^2+z^2}+C,
\end{multline}
where we have introduced  the spectral density of bath $J(z)=2\pi\sum_k\delta(z-\varepsilon_k)$, and $C$ is some unimportant constant  choosen from the condition $F(0)=0$. CL environment corresponds to $J_{CL}(z)=z/(2\pi)$.

\paragraph{Operator of angular momentum.} Many-body representation of the initial problem, introduced above, allow clear and useful interpretation. From the commutation relations it is evident that operator $\hat L\equiv e^{-i\hat\theta}$ increases and operator $\hat L^\dag\equiv e^{i\hat\theta}$ decreases the angular momentum of particle by $1$. On the same time in Appendix A it is shown that operators $\hat a_k$ and $\hat b_k$ can be associated  with two-dimensional oscillator having frequency $\varepsilon_k$. In addition, it is demonstrated, that operators $S_k^\dag\equiv \hat a_k^\dag+\hat b_k\ (\hat S_k\equiv \hat a_k+\hat b_k^\dag)$  changes the angular momentum of given two-dimensional oscillator by $1(-1)$.  In terms of new operators Hamiltonian has the form
\begin{multline}
\hat H_f=\frac{(\hat p-\phi_x)^2}{2mR^2}+\sum\limits_k \varepsilon_k\left(\hat a^\dag_k\hat a_k+\hat b^\dag_k\hat b_k\right)\\+\sqrt{g}\hat L^\dag \sum_k\hat S_k
+\sqrt{g}\hat L\sum_k\hat S_k^\dag.
\label{fictl}
\end{multline}
One can see that interaction term is simply the angular momentum exchange between the particle and the environment. It means that angular momentum of the whole system conserves and operator of total angular momentum
\begin{equation}
\hat M=\hat p+\sum\limits_k\left(\hat a^\dag_k\hat a_k-\hat b^\dag_k\hat b_k\right)
\end{equation}
commutes with Hamiltonian $[\hat H_f,\hat M]=0$. From the fact that eigenvalues of $\hat p$ are integers, as well as eigenvalues of $\hat a_k^\dag\hat a_k$ and $\hat b^\dag_k\hat b_k$, it follows that eigenvalues of $\hat M$ are also integers. Therefore any non-degenerate  eigenstate of Hamiltonian has definite integer angular momentum, i.e. it is  also the eigenstate of $\hat M$ with integer eigenvalue.  Among   others it is true for the ground state of the system means that the expectation value of the total angular momentum in the limit of zero temperature is integer, so  $p_{eff}\equiv\langle \hat M\rangle_{T\to0}$ is quantized and equals to $0,\pm1,\pm2,...$. Note, that in the non-interacting  limit ($g\to0$) $p_{eff}$ is simply the angular momentum of the particle. Below we will show that quantization of $p_{eff}$ is precisely the same thing as in the work of  Burmistrov and Pruisken \cite{BP2008,BP2010}. But we've proved not only the existence of macroscopically quantized observable, but the presence of the additional conservation law  in  the system. It generates a lot of new relations. For example it leads to the absence of the fluctuations of $\hat M$ in the zero temperature limit despite the  fact that fluctuations of particle's angular momentum are finite \cite{SZ2010,SZ2011}. 

\paragraph{Observables and generating function.}  Let us show that $p_{eff}$ is the same observable, as  an ``effective charge'' of  Burmistrov and Pruisken. In order to calculate the average of the total angular momentum, as well as its fluctuations it is useful to introduce generation function by the relation
\begin{equation}
G(\chi)=-\log\left(\frac{\tr(e^{-\beta\hat H_f+i\chi \hat M})}{\tr(e^{-\beta\hat H_f})}\right). 
\end{equation}
This function  is $2\pi$ - periodic and its derivatives are related to the cumulants of  $\hat M$.
\begin{equation}
G'(0)=-i\langle\hat M\rangle\quad G''(0)=\langle\hat M^2\rangle-\langle\hat M\rangle^2,
\end{equation}
where averaging $\langle...\rangle=\tr(...e^{-\beta\hat H_f})/\tr(.e^{-\beta\hat H_f})$ is performed over equilibrium thermal state of the whole system. Note, that the combination $\hat H_f+iT\chi\hat M$ has the same form as $\hat H_f$ but with slightly modified parameters. It means that one can perform calculations along the same line as before. The result is
\begin{equation}
G(\chi)=G_b(\chi)+\log\left(\frac{\mathcal Z}{\mathcal Z_\chi}\right),
\end{equation}
where 
\begin{equation}
G_b(\chi)=\int\limits_0^\infty \frac{dz}{2\pi} J(z)\log\left(\frac{\cosh(\beta z)-\cos(\chi)}{\cosh(\beta z)-1}\right)
\label{Gb}
\end{equation}
is the contribution from the environment, $\mathcal Z$ is given by equation (\ref{partf}), and $\mathcal Z_\chi$ is given by the same expression, but with modified action
\begin{multline}
 S_\chi[\theta]=-i\phi_x\chi+\frac{mR^2}{2}\int\limits_0^\beta(\dot\theta(\tau)+\chi T)^2d\tau\\+\frac{g}{4}\int\limits_0^\beta d\tau\int\limits_0^\beta d\tau'\alpha_\chi(\tau-\tau')e^{i\theta(\tau)-i\theta(\tau')}. 
 \label{modact}
\end{multline}
The modified kernel $\alpha_\chi(\tau)=\frac{T}{\pi}\sum_n F_\chi(\omega_n)e^{-i\omega_n\tau}$ is the following
\begin{equation}
F_\chi(\omega_n)=-8\pi\int\limits_0^\infty \frac{dz}{2\pi} \frac{zJ(z)}{(\omega_n+\chi T)^2+z^2}+C.
\end{equation}
These expressions are general. Let us derive expression for $\langle\hat M\rangle$ by taking the derivative over $\chi$. The result is
\begin{multline}
\langle \hat M \rangle=\phi_x+imR^2T\int\limits_0^\beta d\tau\langle\dot\theta(\tau)\rangle_{\theta}\\
+\frac{igT}{4}\int\limits_0^\beta d\tau\int\limits_0^\beta d\tau'\mu(\tau-\tau')\langle e^{i\theta(\tau)-i\theta(\tau')}\rangle_\theta.
\end{multline}
where $\mu(\tau)=\frac{T}{\pi}\sum_n F'(\omega_n)e^{-i\omega_n\tau}$ and by $\langle...\rangle_\theta$ we denote averaging over all particle trajectories as in equation ({\ref{partf}}). Introducing correlator
\begin{equation}
D_k(\omega_n)=\int\limits_0^\beta d\tau e^{i\omega_n\tau}\langle e^{-ik\theta(\tau)+ik\theta(0)}\rangle_\theta,
\end{equation}
and using the fact that $\langle e^{-ik\theta(\tau)+ik\theta(0)}\rangle_\theta$ is purely real, which is followed from  the $\theta\to-\theta$ invariance, one can rewrite the average total angular momentum as
\begin{multline}
\langle \hat M \rangle=\phi_x+imR^2\langle\dot\theta(\tau)\rangle_\theta\\-\frac{gT}{2\pi}\sum\limits_{n=1}^\infty F'(\omega_n){\rm Im}D_1(\omega_n).
\end{multline} 
This is nothing but the ``effective charge'' introduced by Burmistrov and Pruisken \cite{BP2008,BP2010}, because in the case of CL bath $F'(\omega_n)=1$ for $n>0$. As we have shown above this quantity is quantized in the zero temperature limit $T\to 0$. Note, that in this limit generating function can be calculated exactly. It is linear $G(\chi)=iM_0\chi$ for $-\pi\leq\chi\leq\pi$ and $M_0$ is an integer number equals to angular momentum in the  ground state. Therefore in the zero temperature limit all cumulants except first one tend to zero.

As  the another example, let us consider the  generating function in the case of CL environment. Evaluating integral in equation (\ref{Gb}) one can obtain
\begin{equation}
 G_b^{CL}(\chi)=\frac{T^2}{4\pi^2}\left(2\zeta(3)-{\rm Li_3}(e^{i\chi})-{\rm Li_3}(e^{-i\chi})\right),
\end{equation}
where ${\rm Li_n}(x)$ is a Polylogarithm function, and $\zeta(x)$ is a Riemann Zeta function. At the same time for this case $F_\chi(\omega_n)=|\omega_n+\chi T|$ and
\begin{equation}
 \alpha_\chi^{CL}(\tau)=-T^2\frac{e^{2\pi i T\tau\left[\frac{\chi}{2\pi}\right]}\left(1+2i\left\{\frac{\chi}{2\pi}\right\}e^{\pi iT\tau}\sin(\pi T\tau)\right)}{\sin^2(\pi T\tau)}
\end{equation}
where by $[...]$ and $\{...\}$ we denote integer and fractional parts. From these expressions one can see that $G''(\chi)$ is singular for $\chi\to 0$ means that  the dispersion of $\hat M$ is infinite at the any nonzero temperature.  This nontrivial fact follows from the property of CL environment. It contains a large number of low-lying eigenstates, so one can easily excite bath to the state with any large angular momentum. 

\paragraph{Discussion.} In this letter we considered only the  particle on the ring interacting with the dissipative environment, but this system is equivalent to the well-known AES description of single-electron devices and therefore all results can be directly applied to that case. In particular we have rigorously proved the existence of the macroscopically quantized observable and shed the light on its physical nature.  It emerged that the phenomenon of quantization in the considered system is based on the two  key points. First one is the existence of rotational symmetry and correspondingly the existence of the conserving quantity which we represented by operator $\hat M$ commuting with Hamiltonian $\hat H_f$. The another ingredient for quantization is the topological structure of the particle configuration space, or, in other words, discreteness of the bare particle angular momenta. In the case of AES model the last condition means the  discreteness of charge.  This two points is rather general and therefore the phenomenon of macroscopic quantization not only the property of AES 
model, but can be observed in a variety of other systems. For example, one can consider particle interacting with  more general dissipative environment represented by action
\begin{equation}
 S_i[\theta]=\frac{g}{4}\sum\limits_{q=1}^\infty \int\limits_0^\beta d\tau\int\limits_0^\beta d\tau'\alpha_q(\tau-\tau') e^{iq\theta(\tau)-iq\theta(\tau')},
\end{equation}
and  derive the corresponding total angular momentum which is
\begin{multline}
\langle \hat M \rangle=\phi_x+imR^2\langle\dot\theta(\tau)\rangle_\theta\\-\frac{gT}{2\pi}\sum\limits_{n=1}^\infty \sum_{q=1}^\infty q F_q'(\omega_n){\rm Im}D_q(\omega_n).
\end{multline} 
Here by definition  $\alpha_q(\tau)=\frac{T}{\pi}\sum_n F_q(\omega_n)e^{-i\omega_n\tau}$. Precisely the same arguments as we used in letter can be employed  for demonstration of  $\langle \hat M \rangle$ quantization in the zero temperature limit. One of the important physical examples is the particle interacting with the dirty electron gas environment which was previously used for the investigation of the interplay between quantum coherence and electron-electron interactions in dirty electronic systems (see \cite{GHZ2003,SZ2011} and references therein). In that case $g=3\pi/(2 k_F l_e)^2$, $F_q(\omega_n)=2l_e\log(R/(ql_e))|\omega_n|/R$, and $1\leq q\leq R/l_e$. Here $k_F$ and $l_e$ are the Fermi wavelength and mean free path of the electrons in the bath.

Among the issue with macroscopic quantization there is another interesting question  which originates from the existence of conserving momentum. How does this conservation law influence on the dynamics of the system? Up to now we don't have the complete answer to this question but what we can say is that during the evolution system not necessary reaches the true thermal equilibrium with density matrix $\hat \rho=e^{-\beta\hat H_f}$. For example, if in the initial state system has the definite total angular momentum then the system will satisfy this condition at the any subsequent time moment. This statement is rather simple and intuitive for the particle on the ring, but is nontrivial for the single-electron devices. It means that there are long-lived non-equilibrium states in that case, which can in principle influence the observables. Beyond the physical questions for that case there are some pure technical problems. Commonly used approaches  employ real-time evolution in order to calculate observables at equilibrium. They assume  that during the time  evolution system reaches its thermal equilibrium state independent from the initial one, but in some cases this is not true. This fact restricts the validity of real-time approach. From our point of view it is necessary to investigate this question and it will be a subject of our future work.

In summary, we have demonstrated that the phenomenon of macroscopic quantization has the clear physical meaning in the system of the particle on the ring interacting with dissipative bath. In that case the quantizing observable is simply the total angular momentum of the whole system. We related this phenomenon with angular momentum conservation due to rotational symmetry of the system and rigorously proved that in the ground state total angular momentum can take only integer values. With the aid of the mapping between considered system and the AES model describing single-electron devices we have showed that the macroscopically quantizing angular momentum is an equivalent  of the ``effective charge'' introduced by Burmistrov and Pruisken for the single-electron box problem. Also we have generalized the obtained results to other systems.

\begin{acknowledgements}
The author is grateful to V. Losyakov, G. Starkov and especially to S. Apenko for valuable discussions.
\end{acknowledgements}

\appendix

\section{Appendix A: Two-dimensional harmonic oscillator.} Aim of this appendix is to rewrite Hamiltonian of two-dimensional harmonic oscillator
\begin{equation}
\hat H=\frac{\hat p_x^2+\hat p_y^2}{2 m}+\frac{m\omega^2(\hat x^2+\hat y^2)}{2}
\end{equation}
through creation and annihilation operators. Here $\hat x$, $\hat y$ are position operators and $\hat p_x$, $\hat p_y$ are corresponding momenta operators.   The desired representation is given by the following relations
\begin{equation}
\hat a=\frac{1}{2\sqrt{m\omega}}\left( \hat p_x+i\hat p_y-im\omega\hat x+m\omega\hat y\right),
\end{equation}
\begin{equation}
\hat b=\frac{1}{2\sqrt{m\omega}}\left( \hat p_x-i\hat p_y-im\omega\hat x-m\omega\hat y\right).
\end{equation}
One can check that introduced operators satisfy bosonic commutation relations and  in this representation Hamiltonian simplifies to $\hat H=\omega(\hat a^\dag\hat a+\hat b^\dag\hat b+1)$. On the same time angular momentum operator $\hat J=\hat y\hat p_x-\hat x\hat p_y$ might be rewritten as $\hat J=\hat a^\dag\hat a-\hat b^\dag\hat b$. From the bosonic commutation relation one can show that operator $\hat S=\hat a+\hat b^\dag$ has the following property
\begin{equation}
 \hat S(\hat J-1)=\hat J\hat S \qquad \hat S^\dag(\hat J+1)=\hat J\hat S^\dag
\end{equation}
means that operator $\hat S$ $(\hat S^\dag)$ decreases (increases) angular momentum of oscillator by one.

\bibliography{biblio}{}

\begin{thebibliography}{21}%
\makeatletter
\providecommand \@ifxundefined [1]{%
 \@ifx{#1\undefined}
}%
\providecommand \@ifnum [1]{%
 \ifnum #1\expandafter \@firstoftwo
 \else \expandafter \@secondoftwo
 \fi
}%
\providecommand \@ifx [1]{%
 \ifx #1\expandafter \@firstoftwo
 \else \expandafter \@secondoftwo
 \fi
}%
\providecommand \natexlab [1]{#1}%
\providecommand \enquote  [1]{``#1''}%
\providecommand \bibnamefont  [1]{#1}%
\providecommand \bibfnamefont [1]{#1}%
\providecommand \citenamefont [1]{#1}%
\providecommand \href@noop [0]{\@secondoftwo}%
\providecommand \href [0]{\begingroup \@sanitize@url \@href}%
\providecommand \@href[1]{\@@startlink{#1}\@@href}%
\providecommand \@@href[1]{\endgroup#1\@@endlink}%
\providecommand \@sanitize@url [0]{\catcode `\\12\catcode `\$12\catcode
  `\&12\catcode `\#12\catcode `\^12\catcode `\_12\catcode `\%12\relax}%
\providecommand \@@startlink[1]{}%
\providecommand \@@endlink[0]{}%
\providecommand \url  [0]{\begingroup\@sanitize@url \@url }%
\providecommand \@url [1]{\endgroup\@href {#1}{\urlprefix }}%
\providecommand \urlprefix  [0]{URL }%
\providecommand \Eprint [0]{\href }%
\providecommand \doibase [0]{http://dx.doi.org/}%
\providecommand \selectlanguage [0]{\@gobble}%
\providecommand \bibinfo  [0]{\@secondoftwo}%
\providecommand \bibfield  [0]{\@secondoftwo}%
\providecommand \translation [1]{[#1]}%
\providecommand \BibitemOpen [0]{}%
\providecommand \bibitemStop [0]{}%
\providecommand \bibitemNoStop [0]{.\EOS\space}%
\providecommand \EOS [0]{\spacefactor3000\relax}%
\providecommand \BibitemShut  [1]{\csname bibitem#1\endcsname}%
\let\auto@bib@innerbib\@empty
\bibitem [{\citenamefont {Sch{\"o}n}\ and\ \citenamefont
  {Zaikin}(1990)}]{ShZ1990}%
  \BibitemOpen
  \bibfield  {author} {\bibinfo {author} {\bibfnamefont {G.}~\bibnamefont
  {Sch{\"o}n}}\ and\ \bibinfo {author} {\bibfnamefont {A.~D.}\ \bibnamefont
  {Zaikin}},\ }\href@noop {} {\bibfield  {journal} {\bibinfo  {journal} {Phys.
  Rep.}\ }\textbf {\bibinfo {volume} {198}},\ \bibinfo {pages} {237} (\bibinfo
  {year} {1990})}\BibitemShut {NoStop}%
\bibitem [{\citenamefont {Golubev}\ and\ \citenamefont
  {Zaikin}(1992)}]{GZ1992}%
  \BibitemOpen
  \bibfield  {author} {\bibinfo {author} {\bibfnamefont {D.~S.}\ \bibnamefont
  {Golubev}}\ and\ \bibinfo {author} {\bibfnamefont {A.~D.}\ \bibnamefont
  {Zaikin}},\ }\href@noop {} {\bibfield  {journal} {\bibinfo  {journal} {Phys.
  Rev. B}\ }\textbf {\bibinfo {volume} {46}},\ \bibinfo {pages} {10903}
  (\bibinfo {year} {1992})}\BibitemShut {NoStop}%
\bibitem [{\citenamefont {Golubev}\ and\ \citenamefont
  {Zaikin}(1994)}]{GZ1994}%
  \BibitemOpen
  \bibfield  {author} {\bibinfo {author} {\bibfnamefont {D.~S.}\ \bibnamefont
  {Golubev}}\ and\ \bibinfo {author} {\bibfnamefont {A.~D.}\ \bibnamefont
  {Zaikin}},\ }\href@noop {} {\bibfield  {journal} {\bibinfo  {journal} {Phys.
  Rev. B}\ }\textbf {\bibinfo {volume} {50}},\ \bibinfo {pages} {8736}
  (\bibinfo {year} {1994})}\BibitemShut {NoStop}%
\bibitem [{\citenamefont {K{\"o}nig}\ \emph {et~al.}(1995)\citenamefont
  {K{\"o}nig}, \citenamefont {Schoeller},\ and\ \citenamefont
  {Sch{\"o}n}}]{KSS1995}%
  \BibitemOpen
  \bibfield  {author} {\bibinfo {author} {\bibfnamefont {J.}~\bibnamefont
  {K{\"o}nig}}, \bibinfo {author} {\bibfnamefont {H.}~\bibnamefont
  {Schoeller}}, \ and\ \bibinfo {author} {\bibfnamefont {G.}~\bibnamefont
  {Sch{\"o}n}},\ }\href@noop {} {\bibfield  {journal} {\bibinfo  {journal}
  {Europhys. Lett.}\ }\textbf {\bibinfo {volume} {31}},\ \bibinfo {pages} {31}
  (\bibinfo {year} {1995})}\BibitemShut {NoStop}%
\bibitem [{\citenamefont {Schoeller}\ and\ \citenamefont
  {Sch{\"o}n}(1994)}]{SS1994}%
  \BibitemOpen
  \bibfield  {author} {\bibinfo {author} {\bibfnamefont {H.}~\bibnamefont
  {Schoeller}}\ and\ \bibinfo {author} {\bibfnamefont {G.}~\bibnamefont
  {Sch{\"o}n}},\ }\href@noop {} {\bibfield  {journal} {\bibinfo  {journal}
  {Phys. Rev. B}\ }\textbf {\bibinfo {volume} {50}},\ \bibinfo {pages} {18436}
  (\bibinfo {year} {1994})}\BibitemShut {NoStop}%
\bibitem [{\citenamefont {Golubev}\ \emph {et~al.}(1997)\citenamefont
  {Golubev}, \citenamefont {K{\"o}nig}, \citenamefont {Schoeller},
  \citenamefont {Sch{\"o}n},\ and\ \citenamefont {Zaikin}}]{GKSShZ1997}%
  \BibitemOpen
  \bibfield  {author} {\bibinfo {author} {\bibfnamefont {D.~S.}\ \bibnamefont
  {Golubev}}, \bibinfo {author} {\bibfnamefont {J.}~\bibnamefont {K{\"o}nig}},
  \bibinfo {author} {\bibfnamefont {H.}~\bibnamefont {Schoeller}}, \bibinfo
  {author} {\bibfnamefont {G.}~\bibnamefont {Sch{\"o}n}}, \ and\ \bibinfo
  {author} {\bibfnamefont {A.~D.}\ \bibnamefont {Zaikin}},\ }\href@noop {}
  {\bibfield  {journal} {\bibinfo  {journal} {Phys. Rev. B}\ }\textbf {\bibinfo
  {volume} {56}},\ \bibinfo {pages} {15782} (\bibinfo {year}
  {1997})}\BibitemShut {NoStop}%
\bibitem [{\citenamefont {Eckern}\ \emph {et~al.}(1982)\citenamefont {Eckern},
  \citenamefont {Sch{\"o}n},\ and\ \citenamefont {Ambegaokar}}]{AES1982}%
  \BibitemOpen
  \bibfield  {author} {\bibinfo {author} {\bibfnamefont {U.}~\bibnamefont
  {Eckern}}, \bibinfo {author} {\bibfnamefont {G.}~\bibnamefont {Sch{\"o}n}}, \
  and\ \bibinfo {author} {\bibfnamefont {V.}~\bibnamefont {Ambegaokar}},\
  }\href@noop {} {\bibfield  {journal} {\bibinfo  {journal} {Phys. Rev. Lett.}\
  }\textbf {\bibinfo {volume} {48}},\ \bibinfo {pages} {1745} (\bibinfo {year}
  {1982})}\BibitemShut {NoStop}%
\bibitem [{\citenamefont {Eckern}\ \emph {et~al.}(1984)\citenamefont {Eckern},
  \citenamefont {Sch{\"o}n},\ and\ \citenamefont {Ambegaokar}}]{AES1984}%
  \BibitemOpen
  \bibfield  {author} {\bibinfo {author} {\bibfnamefont {U.}~\bibnamefont
  {Eckern}}, \bibinfo {author} {\bibfnamefont {G.}~\bibnamefont {Sch{\"o}n}}, \
  and\ \bibinfo {author} {\bibfnamefont {V.}~\bibnamefont {Ambegaokar}},\
  }\href@noop {} {\bibfield  {journal} {\bibinfo  {journal} {Phys. Rev. B}\
  }\textbf {\bibinfo {volume} {30}},\ \bibinfo {pages} {6419} (\bibinfo {year}
  {1984})}\BibitemShut {NoStop}%
\bibitem [{\citenamefont {Guinea}(2002)}]{Guinea2002}%
  \BibitemOpen
  \bibfield  {author} {\bibinfo {author} {\bibfnamefont {F.}~\bibnamefont
  {Guinea}},\ }\href@noop {} {\bibfield  {journal} {\bibinfo  {journal} {Phys.
  Rev. B}\ }\textbf {\bibinfo {volume} {65}},\ \bibinfo {pages} {205317}
  (\bibinfo {year} {2002})}\BibitemShut {NoStop}%
\bibitem [{\citenamefont {Guinea}(2003)}]{Guinea2003}%
  \BibitemOpen
  \bibfield  {author} {\bibinfo {author} {\bibfnamefont {F.}~\bibnamefont
  {Guinea}},\ }\href@noop {} {\bibfield  {journal} {\bibinfo  {journal} {Phys.
  Rev. B}\ }\textbf {\bibinfo {volume} {67}},\ \bibinfo {pages} {045103}
  (\bibinfo {year} {2003})}\BibitemShut {NoStop}%
\bibitem [{\citenamefont {Golubev}\ \emph {et~al.}(2003)\citenamefont
  {Golubev}, \citenamefont {Herrero},\ and\ \citenamefont {Zaikin}}]{GHZ2003}%
  \BibitemOpen
  \bibfield  {author} {\bibinfo {author} {\bibfnamefont {D.~S.}\ \bibnamefont
  {Golubev}}, \bibinfo {author} {\bibfnamefont {C.~P.}\ \bibnamefont
  {Herrero}}, \ and\ \bibinfo {author} {\bibfnamefont {A.~D.}\ \bibnamefont
  {Zaikin}},\ }\href@noop {} {\bibfield  {journal} {\bibinfo  {journal}
  {Europhys. Lett.}\ }\textbf {\bibinfo {volume} {63}},\ \bibinfo {pages} {426}
  (\bibinfo {year} {2003})}\BibitemShut {NoStop}%
\bibitem [{\citenamefont {Etzioni}\ \emph {et~al.}(2012)\citenamefont
  {Etzioni}, \citenamefont {Horovitz},\ and\ \citenamefont
  {Doussal}}]{EBD2012}%
  \BibitemOpen
  \bibfield  {author} {\bibinfo {author} {\bibfnamefont {Y.}~\bibnamefont
  {Etzioni}}, \bibinfo {author} {\bibfnamefont {B.}~\bibnamefont {Horovitz}}, \
  and\ \bibinfo {author} {\bibfnamefont {P.~L.}\ \bibnamefont {Doussal}},\
  }\href@noop {} {\bibfield  {journal} {\bibinfo  {journal} {Phys. Rev. B}\
  }\textbf {\bibinfo {volume} {86}},\ \bibinfo {pages} {235406} (\bibinfo
  {year} {2012})}\BibitemShut {NoStop}%
\bibitem [{\citenamefont {Semenov}\ and\ \citenamefont
  {Zaikin}(2011)}]{SZ2011}%
  \BibitemOpen
  \bibfield  {author} {\bibinfo {author} {\bibfnamefont {A.~G.}\ \bibnamefont
  {Semenov}}\ and\ \bibinfo {author} {\bibfnamefont {A.~D.}\ \bibnamefont
  {Zaikin}},\ }\href@noop {} {\bibfield  {journal} {\bibinfo  {journal} {Phys.
  Rev. B}\ }\textbf {\bibinfo {volume} {84}},\ \bibinfo {pages} {045416}
  (\bibinfo {year} {2011})}\BibitemShut {NoStop}%
\bibitem [{\citenamefont {Caldeira}\ and\ \citenamefont
  {Leggett}(1981)}]{CL1981}%
  \BibitemOpen
  \bibfield  {author} {\bibinfo {author} {\bibfnamefont {A.~O.}\ \bibnamefont
  {Caldeira}}\ and\ \bibinfo {author} {\bibfnamefont {A.~J.}\ \bibnamefont
  {Leggett}},\ }\href@noop {} {\bibfield  {journal} {\bibinfo  {journal} {Phys.
  Rev. Lett.}\ }\textbf {\bibinfo {volume} {46}},\ \bibinfo {pages} {211}
  (\bibinfo {year} {1981})}\BibitemShut {NoStop}%
\bibitem [{\citenamefont {Herrero}\ \emph {et~al.}(1999)\citenamefont
  {Herrero}, \citenamefont {Sch{\"o}n},\ and\ \citenamefont
  {Zaikin}}]{HSZ1999}%
  \BibitemOpen
  \bibfield  {author} {\bibinfo {author} {\bibfnamefont {C.~P.}\ \bibnamefont
  {Herrero}}, \bibinfo {author} {\bibfnamefont {G.}~\bibnamefont {Sch{\"o}n}},
  \ and\ \bibinfo {author} {\bibfnamefont {A.~D.}\ \bibnamefont {Zaikin}},\
  }\href@noop {} {\bibfield  {journal} {\bibinfo  {journal} {Phys. Rev. B}\
  }\textbf {\bibinfo {volume} {59}},\ \bibinfo {pages} {5728} (\bibinfo {year}
  {1999})}\BibitemShut {NoStop}%
\bibitem [{\citenamefont {Burmistrov}\ and\ \citenamefont
  {Pruisken}(2008)}]{BP2008}%
  \BibitemOpen
  \bibfield  {author} {\bibinfo {author} {\bibfnamefont {I.~S.}\ \bibnamefont
  {Burmistrov}}\ and\ \bibinfo {author} {\bibfnamefont {A.~M.~M.}\ \bibnamefont
  {Pruisken}},\ }\href@noop {} {\bibfield  {journal} {\bibinfo  {journal}
  {Phys. Rev. Lett.}\ }\textbf {\bibinfo {volume} {101}},\ \bibinfo {pages}
  {056801} (\bibinfo {year} {2008})}\BibitemShut {NoStop}%
\bibitem [{\citenamefont {Burmistrov}\ and\ \citenamefont
  {Pruisken}(2010)}]{BP2010}%
  \BibitemOpen
  \bibfield  {author} {\bibinfo {author} {\bibfnamefont {I.~S.}\ \bibnamefont
  {Burmistrov}}\ and\ \bibinfo {author} {\bibfnamefont {A.~M.~M.}\ \bibnamefont
  {Pruisken}},\ }\href@noop {} {\bibfield  {journal} {\bibinfo  {journal}
  {Phys. Rev. B}\ }\textbf {\bibinfo {volume} {81}},\ \bibinfo {pages} {085428}
  (\bibinfo {year} {2010})}\BibitemShut {NoStop}%
\bibitem [{\citenamefont {Bulgadaev}(2006)}]{Bulgadaev2006}%
  \BibitemOpen
  \bibfield  {author} {\bibinfo {author} {\bibfnamefont {S.~A.}\ \bibnamefont
  {Bulgadaev}},\ }\href@noop {} {\bibfield  {journal} {\bibinfo  {journal}
  {JETP Lett.}\ }\textbf {\bibinfo {volume} {83}},\ \bibinfo {pages} {563}
  (\bibinfo {year} {2006})}\BibitemShut {NoStop}%
\bibitem [{\citenamefont {Apenko}(2006)}]{Apenko2006}%
  \BibitemOpen
  \bibfield  {author} {\bibinfo {author} {\bibfnamefont {S.~M.}\ \bibnamefont
  {Apenko}},\ }\href@noop {} {\bibfield  {journal} {\bibinfo  {journal} {Phys.
  Rev. B}\ }\textbf {\bibinfo {volume} {74}},\ \bibinfo {pages} {193311}
  (\bibinfo {year} {2006})}\BibitemShut {NoStop}%
\bibitem [{\citenamefont {Apenko}(2008)}]{Apenko2008}%
  \BibitemOpen
  \bibfield  {author} {\bibinfo {author} {\bibfnamefont {S.~M.}\ \bibnamefont
  {Apenko}},\ }\href@noop {} {\bibfield  {journal} {\bibinfo  {journal} {J.
  Phys. A: Math. Theor.}\ }\textbf {\bibinfo {volume} {41}},\ \bibinfo {pages}
  {315301} (\bibinfo {year} {2008})}\BibitemShut {NoStop}%
\bibitem [{\citenamefont {Semenov}\ and\ \citenamefont
  {Zaikin}(2010)}]{SZ2010}%
  \BibitemOpen
  \bibfield  {author} {\bibinfo {author} {\bibfnamefont {A.~G.}\ \bibnamefont
  {Semenov}}\ and\ \bibinfo {author} {\bibfnamefont {A.~D.}\ \bibnamefont
  {Zaikin}},\ }\href@noop {} {\bibfield  {journal} {\bibinfo  {journal} {J.
  Phys.: Condens. Matter}\ }\textbf {\bibinfo {volume} {22}},\ \bibinfo {pages}
  {485302} (\bibinfo {year} {2010})}\BibitemShut {NoStop}%
\end{thebibliography}%

\end{document}